# Unsupervised Phase Mapping of X-ray Diffraction Data by Nonnegative Matrix Factorization Integrated with Custom Clustering

Valentin Stanev[1,2,3], Velimir V. Vesselinov[4], A. Gilad Kusne[1,5], Graham Antoszewski[6], Ichiro Takeuchi[1,2*], Boian S. Alexandrov[7*]

[1]Department of Materials Science and Engineering, University of Maryland, College Park, MD 20742, USA

[2]Center for Nanophysics and Advanced Materials, University of Maryland, College Park, MD 20742, USA

[3]Joint Quantum Institute, University of Maryland, College Park, MD 20742, USA

[4]Los Alamos National Laboratory, Earth Science Division, Los Alamos, NM 87545, USA

[5]National Institute of Standards and Technology, Gaithersburg, MD 20899, USA

[6]Department of Mathematics, University of Maryland, College Park, MD 20742, USA

[7]Los Alamos National Laboratory, Theoretical Division, Los Alamos, NM 87545, USA

[*]Correspondence and requests for materials should be addressed to:

I.T. ( takeuchi@umd.edu)

and

B.S.A. ( boian@lanl.gov )

**Abstract**

Analyzing large X-ray diffraction (XRD) datasets is a key step in high-throughput mapping of the compositional phase diagrams of combinatorial materials libraries. Optimizing and automating this task can help accelerate the process of discovery of materials with novel and desirable properties. Here, we report a new method for pattern analysis and phase extraction of XRD datasets. The method expands the Nonnegative Matrix Factorization method, which has been used previously to analyze such datasets, by combining it with custom clustering and cross-correlation algorithms. This new method is capable of robust determination of the number of basis patterns present in the data which, in turn, enables straightforward identification of any possible peak-shifted patterns. Peak-shifting arises due to continuous change in the lattice constants as a function of composition, and is ubiquitous in XRD datasets from composition spread libraries. Successful identification of the peak-shifted patterns allows proper quantification and classification of the basis XRD patterns, which is necessary in order to decipher the contribution of each unique single-phase structure to the multi-phase regions. The process can be utilized to determine accurately the compositional phase diagram of a system under study. The presented method is applied to one synthetic and one experimental dataset, and demonstrates robust accuracy and identification abilities.

**INTRODUCTION**

Combinatorial approach to high-throughput experimental materials science has been successfully used to perform rapid mapping of composition-structure-property relationships in many complex systems. Leveraging recently developed fast and reliable synthesis and characterization tools, compositional phase diagrams can be mapped with a high density of data points on a single library wafer.[1-6] These phase diagrams can then be used to directly connect materials composition to desirable physical properties[1,2]. Because combinatorial methods can generate a large amount of data at high speed, advanced data analysis tools are increasingly in demand.

Crucial for understanding the link between composition, structure and property is determining the constituent phases of materials from structural measurements such as X-ray diffraction (XRD). However, the pace at which such data is generated far outstrips our ability to process it and turn it into actionable knowledge[7]. Analyzing each diffraction pattern individually is tedious and time consuming if done by hand, and there is great interest currently in automating and accelerating the process. However, creating rapid and reliable methods for automatic phase determination from XRD data has proven challenging. Various machine learning tools such as clustering and semi-supervised methods have been tested for this application.[8-16] One very promising technique for analyzing XRD data is Nonnegative Matrix Factorization (NMF); it is a powerful and reliable method, which decomposes the observed data into strictly additive mix of relatively few non-negative end members. These end members directly represent the diffraction patterns of the structures present at a given multi-phase region. Due to the ease of interpreting its results, NMF has been used in several unsupervised and semi-supervised systems for decomposing XRD patterns, and has given very encouraging results[11,17,18]. However, some significant challenges to utilizing the full potential of this method remain.

One of the hurdles in analyzing XRD data is the presence of diffraction patterns that correspond to the same structure but with shifted diffraction peaks: alloying or solid solution leads to crystal lattice expansion or contraction, which causes a systematic shift of the peak positions in the X-ray patterns. Thus, due to relatively small changes in crystal lattice parameters created by continuous changes in the composition, compounds with nominally the same structure can produce different diffraction patterns. In the NMF analysts this peak-shifting leads to the appearance of several end members representing the same structure, which significantly complicates the determination of the constituent phases present in the compositional phase diagram[11,14,17,18].

Various data analysis methods, both unsupervised and supervised, have been developed in particular to address the problem of automatic identification of the peak-shifted patterns. Some utilize simple clustering tools but with distance measures more resilient to shifting[18,14]. Other methods rely on more abstract constraint programming concepts, and use powerful but computationally expensive metrics like dynamic time warping[19,11]. Several generalizations of NMF techniques, including shiftNMF and convolutional NMF have been used as well[15]. (We discuss some related methods in Materials and Methods section.) Despite its prevalence and practical importance, this issue has not been fully resolved so far, and most NMF-based unsupervised methods are not able to satisfactory identify peak-shifting in the data.

In this work, we take a different, simpler approach. We present a method that extends the NMF algorithm, by organically combining two procedures. The first procedure expands on the conventional NMF by adding a robust protocol for determination of the number of the end members. The second procedure examines the patterns obtained at the end of first procedure for peak-shifting. It estimates, via cross-correlation analysis, the unique end members of the investigated dataset, combines the peak-shifted copies, and modifies accordingly the abundancies obtained from the NMF analysis. The thus-extracted unique end members represent the constitutive single phases, and their abundancies can be used for determining the compositional phase diagram of the system.

Below, we first introduce the method and then demonstrate its capabilities for distinguishing the peak-shifted XRD patterns and constructing compositional phase diagrams, by applying it to both synthetic and experimental XRD datasets.

Because peak-shifting is ubiquitous in many characterization datasets, such as Fourier transform infrared spectroscopy, Raman spectroscopy, and X-ray photoelectron spectroscopy (where in each technique peak-shifting takes place for different reason), the method presented in this work is applicable not only to XRD datasets, but could be used for analyzing large variety of materials data.

**Non-negative Matrix Factorization with custom clustering: NMFk**

NMF is a well-known unsupervised machine learning method created for parts-based representation[20,21] that has been successfully leveraged for decomposing of mixtures formed by various types of non-negative signals[22]. By enforcing only the non-negativity constraint NMF can decompose large sets of observations into a small set of easily interpretable end members. This technique is especially appealing for analyzing XRD data, as the end members directly represent the diffraction patterns of the different crystal structures, and the corresponding weights reflect the abundance of each structural phase at a given nominal composition.

In the NMF formulation of XRD problem, the $N$ experimentally measured (or generated) diffraction patterns form an observational data matrix $X$; $X \in M_{NM}(\mathcal{R}_+)$, where $\mathcal{R}_+$ denotes the set of real non-negative numbers. At each one of these $N$ patterns, and at each one of the $M$ diffraction angles $2\theta$ (or $q$-vectors) of these patterns, the value of the matrix $X_n(2\theta)$ is formed by a linear mixing of $K$ unique but unknown end members. These end members form the unknown matrix, $W, W \in M_{MK}(\mathcal{R}_+)$, blended by an also unknown mixing matrix, $H, H \in M_{KN}(\mathcal{R}_+)$. The values of $H$ correspond to the contribution of each end member in a given point in composition space. Thus, for a given patterns, $X_n$, and at a given angle, $2\theta$, we can write

$$X_n(2\theta) = \sum_{i=1}^{K} W_k(2\theta) H_{k;n} + \varepsilon_n(2\theta),$$

where $\varepsilon \in M_{NM}(\mathcal{R}_+)$ denotes the potential presence of (also unknown) noise or unbiased errors in the measurements. The goal of NMF algorithm is to retrieve the $K$ original non-negative basis patterns (encoded in $W$) that produced the $N$ measured intensity patterns $X$. Since both factor matrices $W$ and $H$ are unknown, and even their size $K$ (i.e., the number of end members) is unknown the problem is under-determined. NMF can solve such kind of problems by leveraging, for example, the multiplicative update algorithm[21] to minimize the Frobenius norm $\frac{1}{2}||X - W * H||_F^2$ or the Kullback–Leibler (KL) divergence $D(X||W * H)$. Application of conventional NMF has demonstrated significant promise in reducing the time and the effort required to analyze large XRD datasets. However, there are some significant challenges in using this method for completely automating this task.

One of the complications of the classical NMF algorithm is that it requires *a priori* estimate of $K$ - the number of end members. Recently a new protocol called NMFk addressing this limitation has been reported[23]. This protocol complements the classical NMF with a custom semi-supervised clustering and Silhouette[24] statistics, which allows simultaneous identification of the optimal number and shapes of the unknown basis patterns. The NMFk protocol was used to successfully decompose the largest available dataset of human cancer[25] genomes, as well as physical transients[26] and contaminants[27] originating from an unknown number of sources.

NMFk determines the number of the unknown basis patterns based on the most robust and reproducible NMF solution. It explores consecutively all possible numbers of end members $\widetilde{K}$ ($\widetilde{K}$ can go from 1 to N, where N is the total number of individual X-ray patterns), by obtaining sets of a large number of NMF minimization solutions for each $\widetilde{K}$. Note that $\widetilde{K}$ serves to index the different NMF models, and is distinct from $K$, which is fixed, albeit unknown number. Then NMFk leverages a custom clustering using Cosine distance (see Materials and Methods section for details), in order to estimate the robustness of each set of solutions with fixed $\widetilde{K}$ but with different initial guesses. Comparing the quality of the derived clusters (a measure of reproducibility of the extracted end members) and the accuracy of minimization among the sets with various $\widetilde{K}$, NMFk determines the optimal numbers of the end members in the data.

To access the quality of the clusters obtained for each set we use their average Silhouette width. NMFk uses it to measure how good is a particular choice of $\widetilde{K}$ as an estimate for $K$. Specifically, the optimal number of patterns is picked by selecting the value of $\widetilde{K}$ that leads to both (a) an acceptable reconstruction error $R$ of the observation matrix V, where

$$R = \frac{\|X-W*H\|_F}{\|X\|_F},$$

and (b) a high average Silhouette width (i.e., an average Silhouette width close to one).

The combination of these two criteria is easy to understand intuitively. For solutions with $\widetilde{K}$ less than the actual number of patterns ($\widetilde{K} < K$) we expect the clustering to be good (with an average Silhouette width close to 1), because several of the actual patterns could be combined to produce one "super-cluster", however, the reconstruction error will be high, due to the model being too constrained (with too few degrees of freedom), and thus on the under-fitting side. In the opposite limit of over-fitting, when $\widetilde{K} > K$ ($\widetilde{K}$ exceeds the actual number of patterns), the average reconstruction error could be quite small - each solution reconstructs the observation matrix very well - but the solutions will not be well-clustered (with an average Silhouette substantially less than 1), since there is no unique way to reconstruct $X$ with more than the actual number of patterns, and no well-separated clusters will form.

Thus, our best estimate for the true number of end-members K is given by the value of $\widetilde{K}$ that optimizes both of these metrics simultaneously. Finally, after determining $K$, we use the centroids of the $K$ clusters to represent the final robust basis end members.

**Identifying the shifted end members**

When previously NMF has been used to analyze XRD patterns from a combinatorial materials library, it has been observed that it often extracts a set of very similar patterns[17]. These patterns differ only by a small shift in the positions of the main peaks, caused by systematic changes in the compound composition and the corresponding expansion or contraction of the lattice. These peak-shifted patterns represent the same crystal structure, and thus should not be considered separate end members.

Peak-shifting in XRD data is a hallmark of alloying, and is appears in any composition spread library. Its presence, however, significantly complicates the analysis of the data. To resolve this problem, we first determine the number of end members (including the shifted ones) via NMFk, and then apply an additional procedure to identify the shifted patterns among them. Specifically, after the NMFk protocol yields the set of $w_1, w_2, ..., w_K$ end members, we estimate the cross-correlations between each pair of end members, $w_i$ and $w_j$ ($i \neq j$), as a function of all possible shifts. The procedure measures the correlation between $w_i$ and all shifted copies of $w_j$ as a function of the shift values, and identifies the shift at which the correlation of the two data-vectors is highest (i.e., they are "most alike"). To identify the actual peak-shifted patterns, only the pair patterns that obey the following two additional constraints are selected:

a) The size of the shift giving the maximum of the cross-correlation has to be in a limited interval. This condition is dictated by the relatively small range of lattice parameters producing the peak-shifting (typically no larger than 0.005 nm). This corresponds to $2\theta$ interval of about 1° ($\pi/180$ radians) for diffraction with Cu K-alpha line ($\lambda \approx 0.154$ nm). (Peak-shifting larger than 0.005 nm is often an indication of a structural transition, and thus, the XRD patterns with shifted peaks should be correctly identified as representing distinct structural phases[28].)

b) The Pearson correlation coefficient between each two pairs $w_i$ and $w_j$ has to be above some high value (we use 0.97), to make sure that one of them is indeed a shifted copy of the other.

We demonstrate below that based on the cross-correlation analysis constrained by these two criteria the peak-shifted copies of the end members can be successfully identified without human supervision. Thus, our method has the ability to automatically recognize small lattice changes that appear in XRD data analyses.

As a practical aside, note that the constraint (b) for a high value ($\geq 0.97$) of the Pearson correlation between two (shifted) patterns works very well when the level of the noise in the data is small, e. g., for a synthetic or for high-quality experimental data (or data to which some smoothing filter has been applied). However, if there is significant noise in the intensity patterns, some peak-shifted end members that differ only by variations of the noise can be missed by requiring such a high degree of correlation. In this case we can relax the second constraint by reducing the cut-off value to 0.90 or even lower, but add as a third constraint the following criterion:

- c) To consider two basis patterns identical (up to a shift), we require the number of their peaks to be equal. This number can be found by estimating the number of the local maxima in the two spectra, with values exceeding the noise level.

Applying this criterion requires an estimation of the level of the noise in the patterns, which also allows to determine if NMFk extracts noise as a separate basis pattern (see Results section).

**Creating a phase diagram**

Once the basis patterns and their corresponding abundancies have been established, it is relatively straightforward to find the outlines of the phase diagram of the compound, separating single-phase from multi-phase regions. This can be done by mapping the regions consisting of combinations of various crystal phases. Note that the NMFk method does not directly enforce physical constrains like the Gibbs' phase rule (which limits the number of allowed phases at a given composition) or connectivity of the phases in composition space. However, if the method works as expected and extracts sufficiently accurate solutions, its results should only permit minor violations of such rules. Thus, the degrees of conformity with these basic constrains provides an estimate of the accuracies of the extracted end members and especially abundancies, both of which are limited by the data (the presence of noise) as well as the method itself (finding suboptimal solutions).

**RESULTS**

Here, we test the proposed method by applying it to the problem of identifying the basis patterns of one synthetic and one experimental XRD datasets. We demonstrate that in both cases the method can successfully perform an unsupervised analysis of the data, including identifying the peak-shifted patterns, and extracting the corresponding phase mapping.

**Synthetic data**

We apply our method to a synthetic dataset, representing XRD patterns of ternary Aluminum-Lithium-Iron (Al-Li-Fe) oxide system, for which the patterns and their parameters have been theoretically calculated. It was specifically designed as a test and validation tool for various XRD data analysis techniques[29]. The pattern for each phase was simulated as a series of Gaussian peaks with positions and relative intensities derived from XRD database patterns, in this case Inorganic Crystal Structure Database (ICSD) entries. The phase diagram of this system is interesting due to the complete pseudo-binary solubility of $Li(Al,Fe)_5O_8$, as well as alloying on the Al and Fe end-member phases. The alloy composition and phase fraction were determined form the known phase diagram, and alloying-based peak shifting was calculated using linear interpolation (a Vegard's Law model).

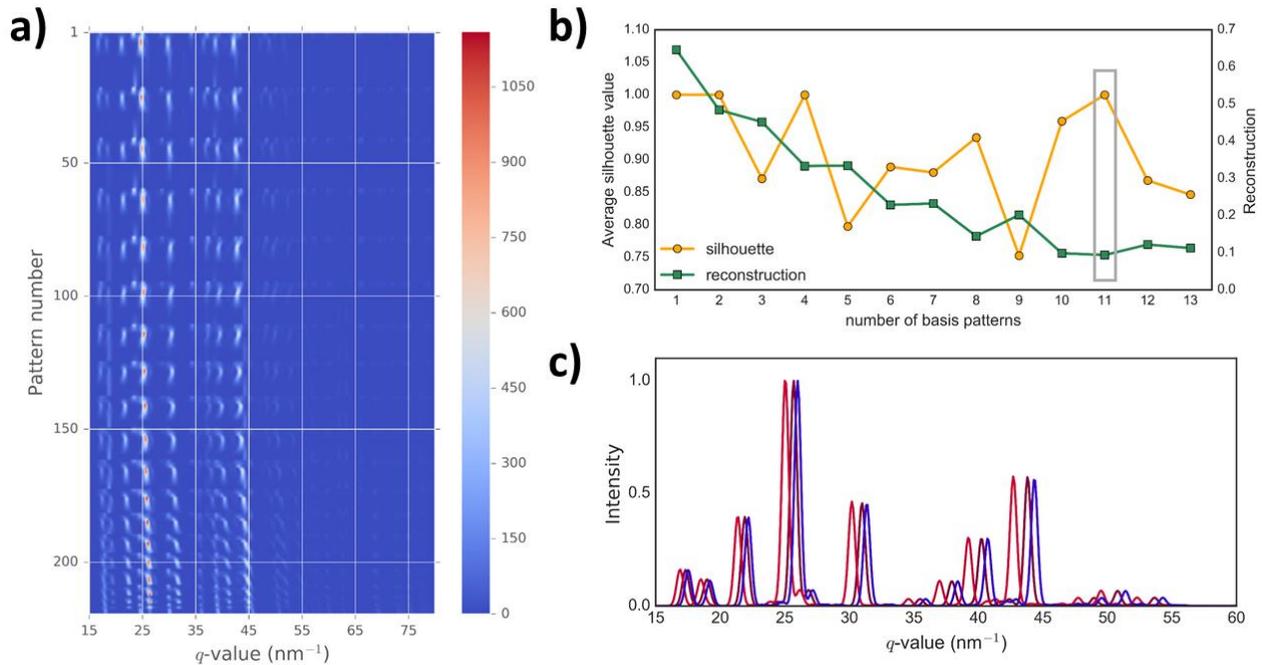

*Figure 1. Al-Li-Fe oxide synthetic data: a) Heat map of the intensity (in arbitrary units) of the Al-Li-Fe oxide system data. b) The average Silhouette values (green curve) obtained by the custom clustering, and error of the NMF reconstruction (yellow curve). To determine the optimal number of the end members in the data, we select a*

The intensity patterns of the system is shown in Fig. 1**a**: the dataset contains 219 diffraction patterns, each characterized at 650 $q$-value points. It includes patterns with various amounts of phase shift (see Fig. 1**c**); note that a lattice constant change of around 0.005 nm corresponds to a peak-shift in $q$-space of less than 3 nm$^{-1}$. The accompanying solution provides the exact end members and the abundancies used to create the dataset at the first place ("ground truth").

The NMFk method reproduces the observed patterns extremely well – only in four cases the cross-correlation between the real and the reconstructed patterns is below 0.85. The exact solution has six basis patterns, but the presence of the peak-shifting complicates their identification significantly. NMFk finds eleven end members - see Fig. 1**b -** some of which are identical up to a phase shift. The shift-detection procedure based on cross-correlation analysis correctly reduced the number of basis patterns to six. As Fig. 2 shows, the remaining end members extracted by our method are very close to the end members used to generate the Al-Li-Fe dataset (for the shifted patterns we show the ones that are the best match).

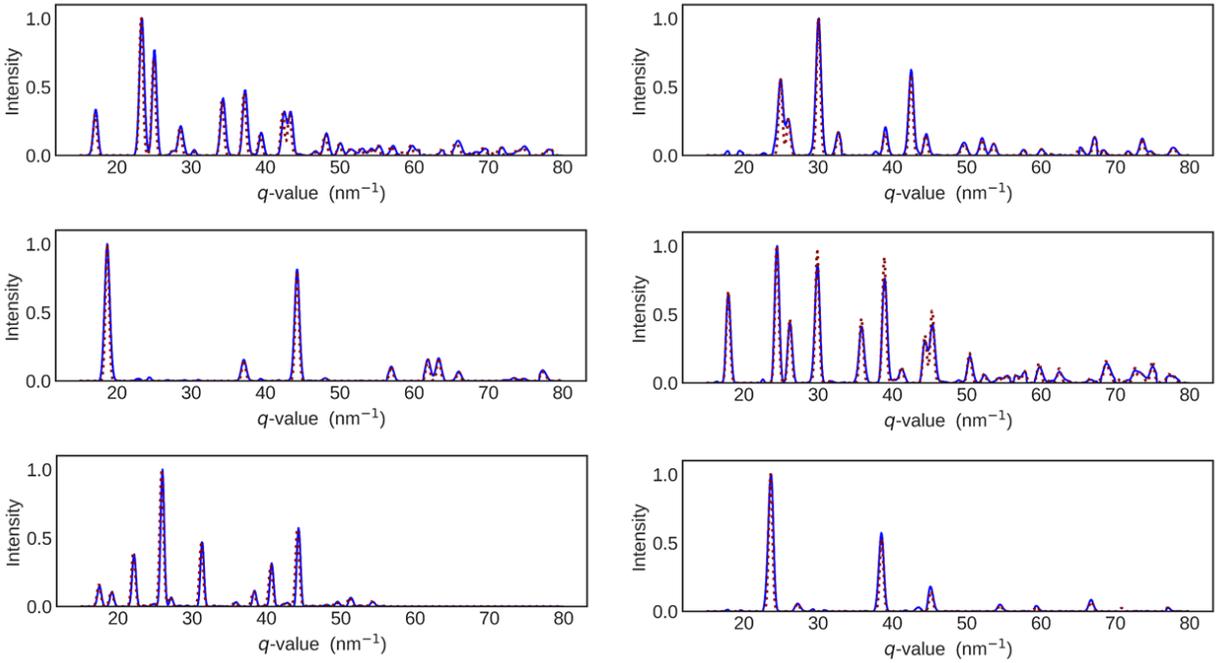

*Figure 2. Al-Li-Fe basis patterns:* *Comparison between the six independent basis patterns as calculated by NMFk (blue solid line) and the end members used to generate the data (dotted red line). The maximum value in each pattern is normalized to 1.*

The matrix *H* which encodes the abundancies of each basis pattern at a given composition is also reconstructed well (see Fig. 3). Based on the extracted abundancies we can see that there are six well-defined regions, which follow very closely the phase regions of the exact solution. There are some violations of the connectivity requirement (each phase should exist in a connected region of composition space), but they are minor and difficult to see in Fig. 3**b**. In Fig. 4**a** similar information about the phases is plotted in different form - at each point the most abundant structure is shown. Again, comparing this with similar map constructed using the exact solution (Fig. 4**b**) demonstrates that the NMFk results are rather accurate.

From the abundancies of the individual phases we can also reconstruct the outlines of the entire structural phase diagram. As a first step, at each composition point we set to zero the contribution to *H* (the mixing matrix) of any phase with abundancy below a certain cut-off. The cut-off is given roughly by the level of the error of the abundancies yielded by NMFk (we have chosen to remove any phase contributing less than 1.2 % of the total at a given point). This removes most (but not all) of the phases violating the connectivity and the Gibbs' phase rule rule (dictating that for a ternary compound there can be no more than three phases at each composition); in addition we remove abundancies violating the connectivity rule. Dividing the composition space into regions of combinations formed by different processes gives us the phase diagram shown in Fig. 4**c**, which is in good agreement with the known phase diagram of the Al-Li-Fe system[30]. It has three regions in the center of the composition space containing combinations of three different phases; the rest contains single phase or two coexisting phases.

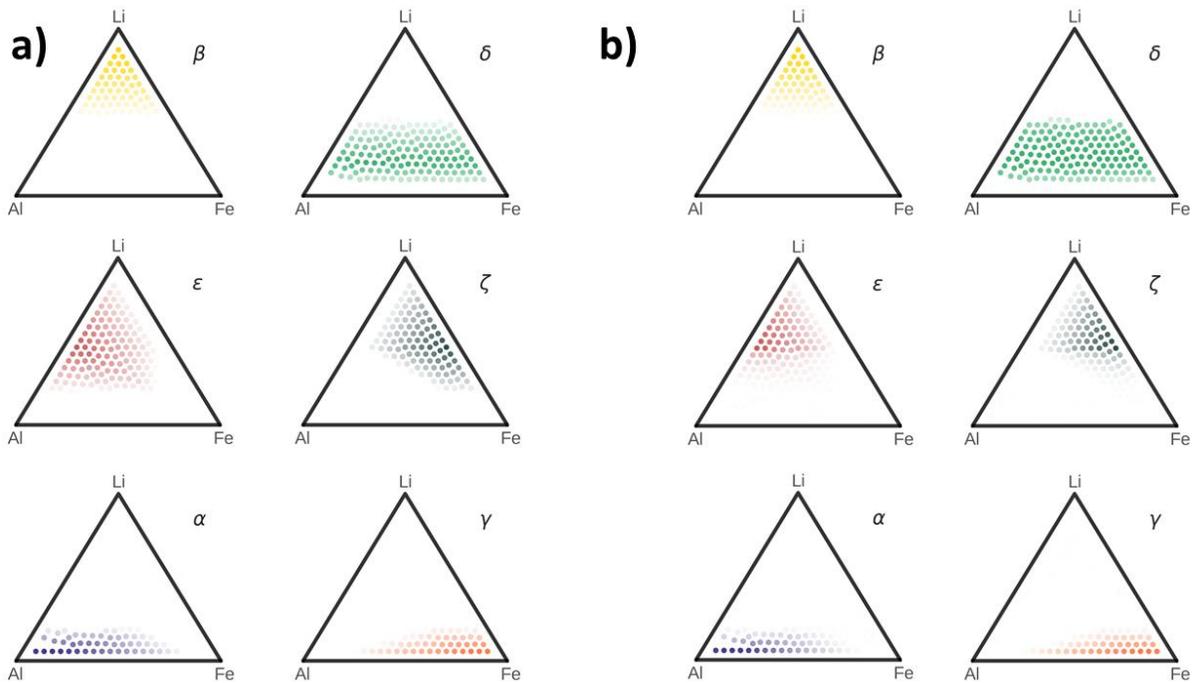

*Figure 3. Phase weights for the Al-Li-Fe oxide data:* *The weights of each unique end member at a given composition. On the left (**a**) is the "ground truth" exact solution (from Ref. [[11]]), and on the right (**b**) are the weights as determined by NMFk method.*

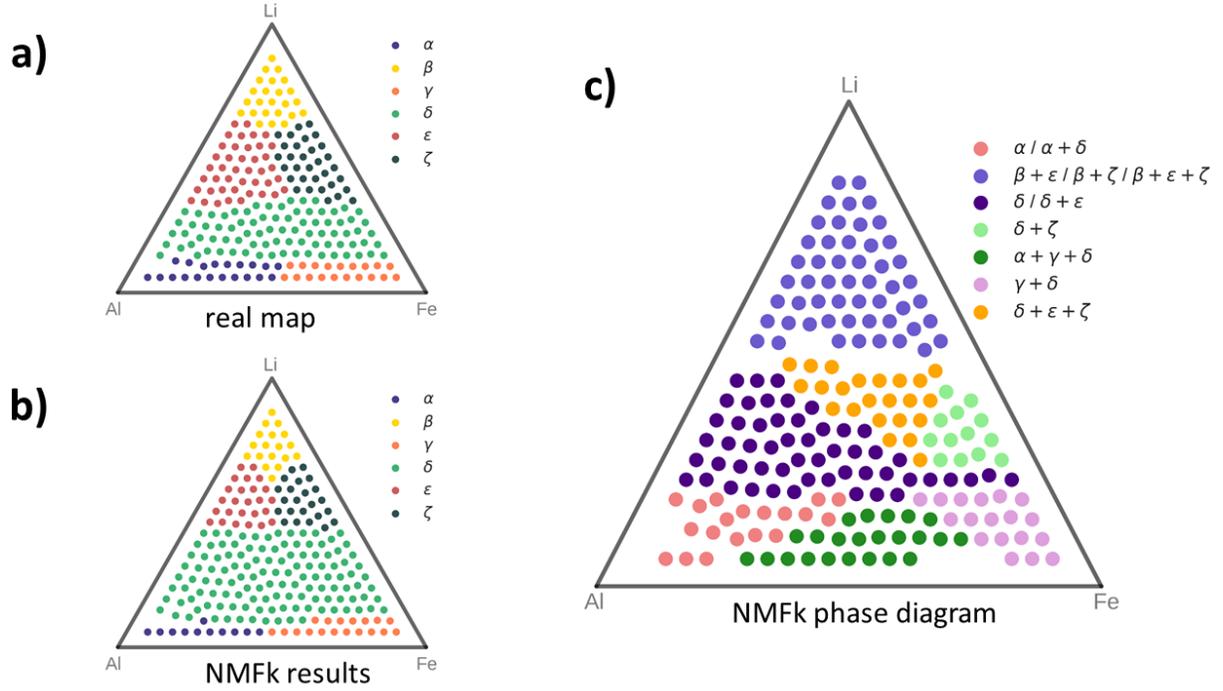

*Figure 4. Phase diagram constructed for the Al-Li-Fe data: The strongest phase at each composition as given by (**a**) the exact solution and (**b**) the NMFk results. **c**) Simplified phase diagram representing the mixing of the different phases.*

**Experimental data for FeGaPd composition spread**

To demonstrate the ability of the proposed method to deal with real (unavoidably noisy and imperfect) experimental data we apply it to the XRD dataset for Fe–Ga–Pd ternary system taken from Ref [17]. This system has been studied extensively, due to the fact that both Fe–Ga and Fe–Pd binary phase diagrams contain compositions with unusual magnetic properties: Fe–Ga system exhibiting large magnetostriction for Ga content between 20 % and 30 %, while $Fe_{70}Pd_{30}$ is a ferromagnetic shape memory. Both Ga and Pd form solid solutions when they are substituted into the Fe lattice up to about 25 %, and could be introduced without disturbing the original Fe crystal structure.

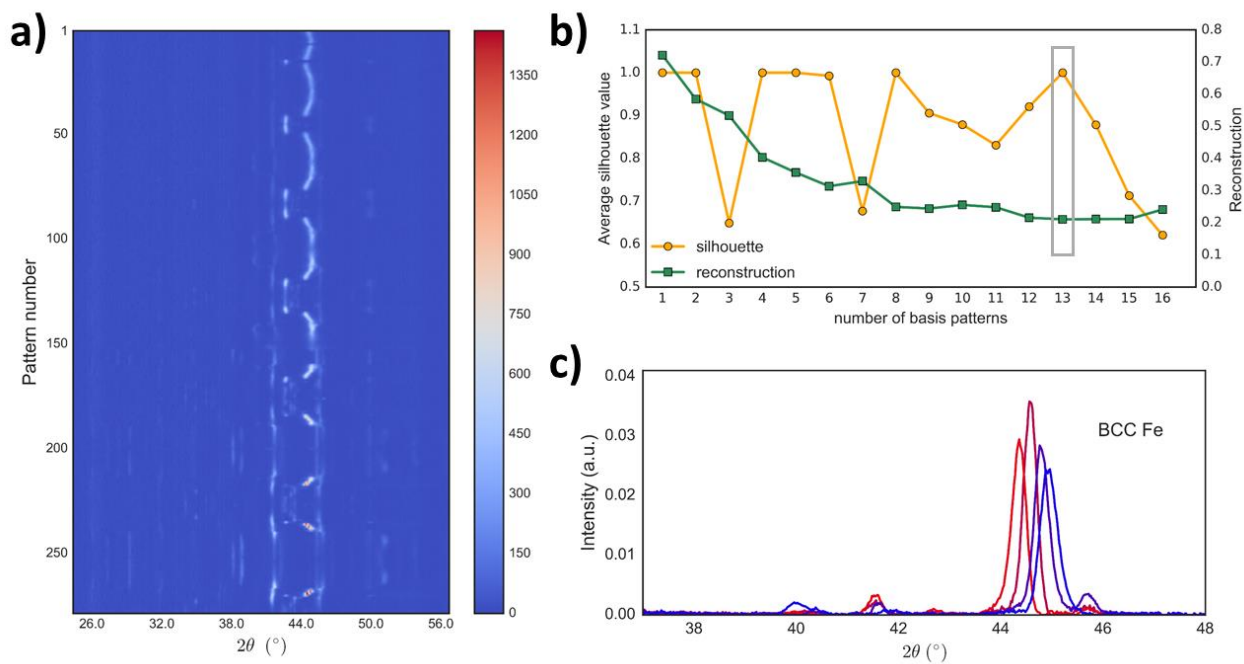

*Figure 5. Fe-Ga-Pd data: a) Heat map of the intensity of the XRD data. b) The average Silhouette values (green curve) obtained by the custom clustering, and error of the NMF reconstruction (yellow curve). The optimal number of basis patterns in this data is 13 (shown by the grey rectangle). c) Four basic end members of the Fe-Ga-Pd system, obtained by NMFk, and differing only by a peak-shift. These patterns represent BCC Fe (α-Fe) structure.*

The data came from a combinatorial library synthesized by co-sputtering Fe, $Fe_2Ga_3$ and Pd targets onto a Si wafer. The library was characterized by 278 patterns via XRD, using the Cu K-alpha line. The entire dataset is visualized in Fig. **5a**; it has been studied and used extensively to test different data analysis methods[9,13,15].

We analyze the entire dataset as observed, without any additional preprocessing (e.g., filtering or noise reduction) apart from background subtraction. Due to the presence of noise, the reconstruction achieved by NMFk is significantly worse overall: in 51 cases (around 18 % of the entire dataset) the cross-correlation between the observed and the reconstructed patterns is below 0.85. NMFk produces 13 end members (see Fig. 5**b**). The peak-shifting detection procedure yields that four of the basic patterns are identical up to a peak shift (Fig 5**c**). (The small peak around 40° and 41.5º have been designated as noise by the algorithm, although they could be traces of other phases.) These patterns have been identified before, and represent BCC Fe (α-Fe) structure with slight difference in the lattice parameters. Plotting the weights of the peak-shifted patterns in composition space (shown in Fig. 6) and concurrently analyzing the shifts themselves, we can understand the changes in the Fe lattice with the inclusion of Pa and Ga atoms. Based on this, we can say that in the BCC Fe phase the lattice expands more (peaks move to smaller diffraction angles) when the Ga content increases relative to Pd. This effect is expected, given the larger ionic radius of Ga.

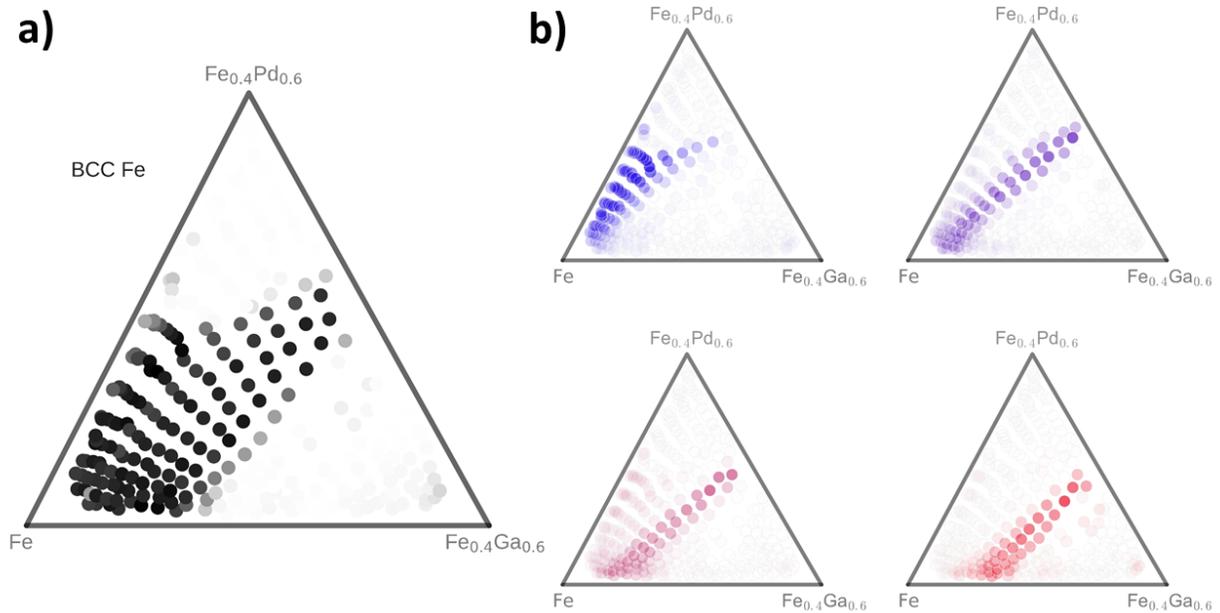

*Figure 6. Peak shifting in the Fe-Ga-Pd system: **a**) The combined weights for the four basis patterns corresponding to the BCC Fe structure. **b**) The weight of each peak-shifted pattern. The evolution of the structure can be deduced from the distribution of the weights and the relative position of the peaks (the colors are matched with those of the patters in Fig. 5c).*

From the remaining nine basis patterns, two have been designated as a noise, due to their very low overall intensity and absence of any peaks above the noise level. Based on the analysis in Ref.[17] we have identified the phases the remaining seven patterns represent (shown in Fig 7). The phase map for the eight basis patterns is shown in Fig. 8**a**. There are some clear violations of the connectivity constrain and the Gibbs' rule. We again remove abundancies contributing less that some fixed value (we use 8 %) to each composition point. Note that this cut-off is much higher than the one used for the synthetic dataset – consequence of the presence of noise in the data.

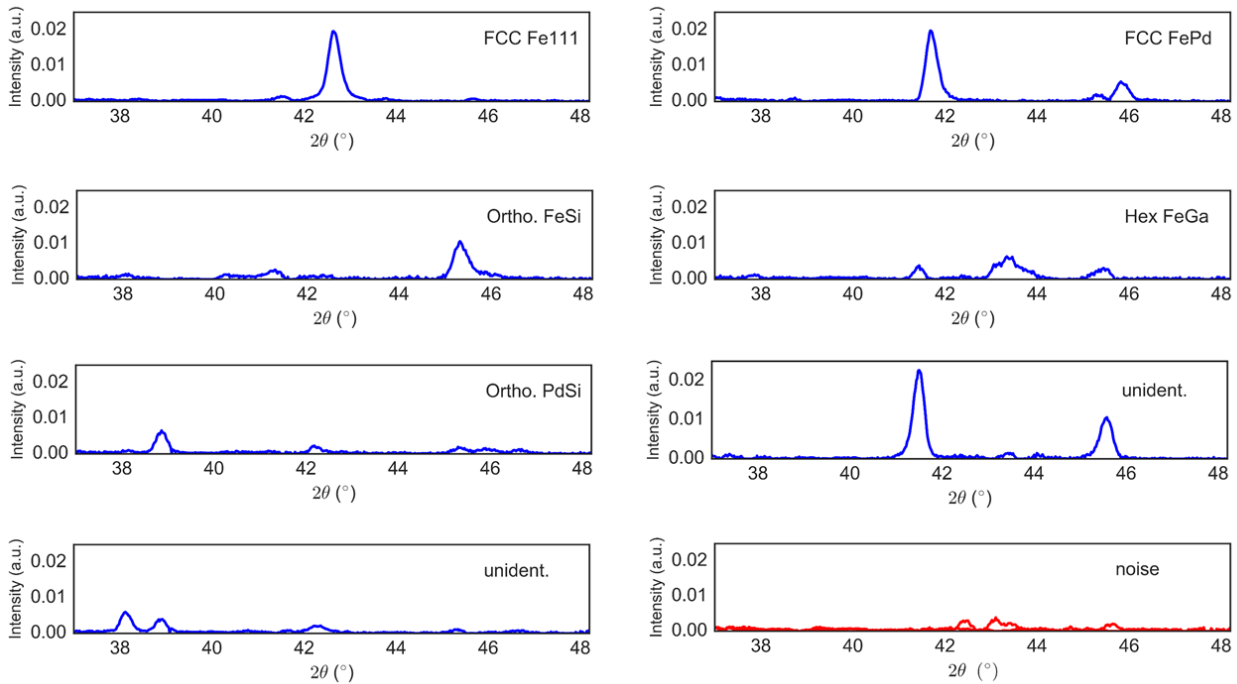

*Figure 7. Basis patterns of the Fe-Ga-Pd system: The end members for Fe-Ga-Pd system as calculated by the method (excluding the peak-shifted patterns corresponding to BCC Fe). The identification is based on Ref.[17]. Two of the patterns in the figure are unidentified and two others represent noise, i.e., they have no significant peaks (based on criterion (c) from page 10).*

At each point we also keep only the three most abundant phases, thus enforcing the Gibbs' rule. Some of the most common combinations formed by the remaining phases give us the phase diagram shown in Fig. 8**a**. The general outlines, including the large α-Fe-dominated central region, are in good agreement with that of other suggested phase diagrams of the Fe-Ga-Pd system. Our phase diagram is more detailed and complicated than the one in Ref. [13], which does not include unidentified phases, but is somewhat simpler that the one from Ref. [9], which contains three of those (versus two in the our analysis).

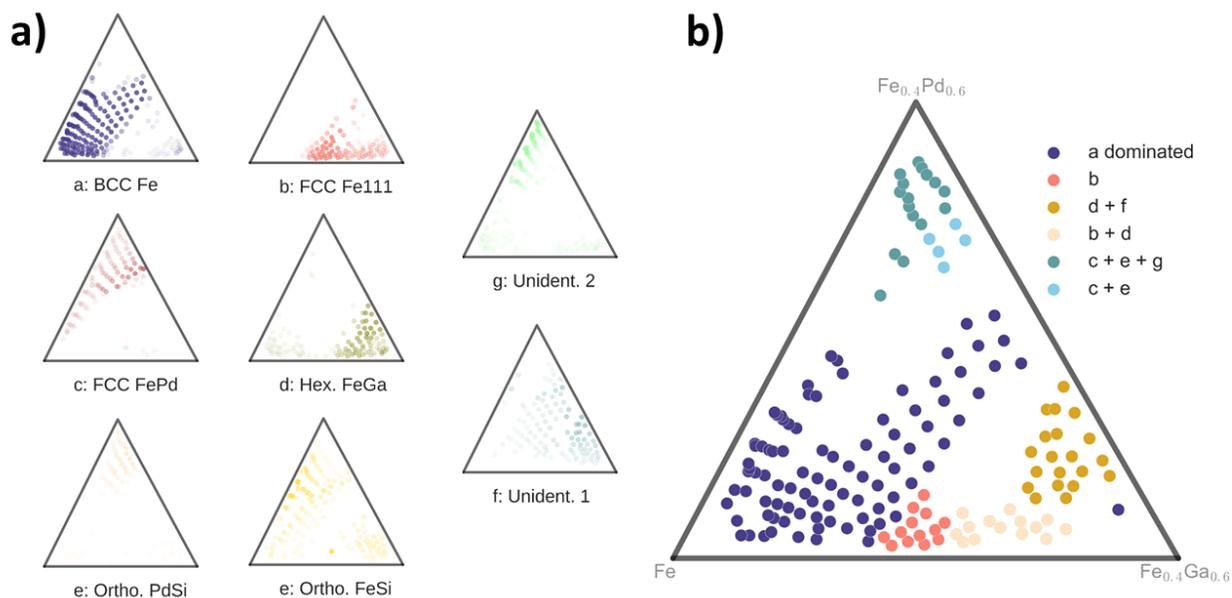

*Figure 8. Structural phases and constructed phase diagram for the Fe-Ga-Pd data: a) The abundancies of the structures recovered by NMFk (noise excluded). b) Simplified phase diagram showing some of the major phases in the Fe-Ga-Pd phase diagram.*

**DISCUSSION**

NMF has shown great promise for the task of analyzing large volumes of XRD measurements. Its simplicity and ease-of interpretability offer great advantages, and several systems relying on this method were created and tested successfully on large datasets. However, some aspects of XRD data still present significant challenges, and dealing with them often requires human intervention. One key problem is determining the number of basis patterns in the data. Another is peak-shifting of X-ray patterns – a common consequence of lattice changes caused by alloying. Here we presented a method that addresses both of these problems. First, it extends the conventional NMF algorithm by leveraging a robust protocol for determining the number of NMF-extracted end members. A second procedure examines the obtained basis patterns for peak-shifting.

By applying the method to both real and synthetic datasets we demonstrated that it can be used for extracting the basis patterns, followed by a quick identification of the peak-shifted ones. We also show that the unique end members and their abundancies can be used for determining the outlines of the compositional phase diagram.

One advantage of our approach of is its simplicity and modularity – it only requires running (can be in parallel) a moderate number of regular NMF minimizations; the computational cost of the added procedures is negligible. Various types of NMF (e.g., sparse NMF[31] or graph regularized NMF[32] that enforce specific constrains, like the Gibbs' phase rule) can be used instead, thus allowing a greater flexibility in analyzing the data. Although at the moment implementing the proposed method still relies on some human input, in principle there are no obstacles to automating all its steps, turning it into an entirely automated system for XRD data analysis.

**MATERIALS AND METHODS**

**NMF minimization algorithm**

Here we leveraged the multiplicative algorithm[21] based on Kullback–Leibler divergence (also called relative entropy)[33], as well as the block coordinate descent algorithm[34], based on Frobenius norm. We did not observe any significant difference between the results obtained via these two algorithms.

**NMFk clustering algorithm**

NMFk creates up to an N sets of minimizations (called NMF runs), one for each possible number $\widetilde{K}$ of original patterns. In each of these runs, $P$ solutions (e.g., $P = 100$) of the NMF minimization, for a fixed number of patterns $\widetilde{K}$, are derived. Thus, each run results in a set of solutions $U_{\widetilde{K}}$:

$$U_{\widetilde{K}} = \{W_{\widetilde{K}}^1, H_{\widetilde{K}}^1; W_{\widetilde{K}}^2, H_{\widetilde{K}}^2; \ldots; W_{\widetilde{K}}^P, H_{\widetilde{K}}^P\},$$

where each of these "tuples" represents a distinct solution for the nominally same NMF minimization, the difference stemming from to the different (random) initial guesses. Next, NMFk performs a custom clustering, assigning the $\widetilde{K}$ columns of each $W_{\widetilde{K}}^i$ of all $P$ solutions to one of the $\widetilde{K}$ clusters, representing $\widetilde{K}$ possible basic patterns. This custom clustering is similar to k-means clustering, but holds the number of elements in each of the clusters equal. For example, with $P = 100$ each one of the $\widetilde{K}$ identified clusters has to contain exactly 100 solutions. This condition has to be enforced since each minimization solution (specified by a given ($W_{\widetilde{K}}^i, H_{\widetilde{K}}^i$) tuple) contributes only one solution for each end member, and accordingly has to supply exactly one element to each cluster. During the clustering, the similarity between patterns is measured using the Cosine distance.

**Complete phase diagrams**

In the main text we presented somewhat simplified versions of the phase diagrams of the Al-Li-Fe and Fe-Ga-Pd systems. For completeness here we show more detailed ones.

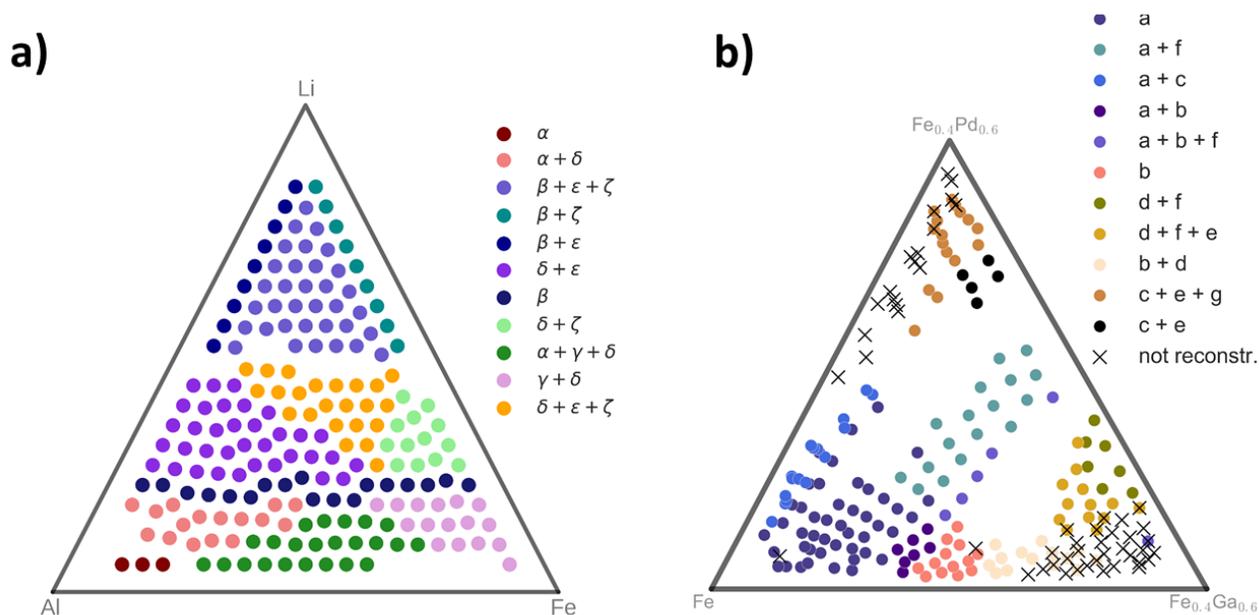

*Figure 9. Detailed phase diagrams for the datasets: **a**) The Al-Li-Fe oxide system phase diagram. **b**) The Fe-Ga-Pd phase diagram.*

**Other algorithms dealing with peak shifting**

There have been several methods proposed specifically to identify peak shifting in various data sets. Several of them have been applied to XRD datasets. Some of these methods (like dynamic time warping) come at a high computational cost, resulting in rather long solution times, and at the moment are of limited practical use.

We have also explored another modification of NMF; shiftNMF[35,36] was developed in the field of signal processing, specifically to look for delays/shifts in signals detected at different locations. shiftNMF seeks to identify the shifts of basis end members at each sample point in the material by introducing the matrix $T$ of peak-shifting parameters. An individual element $T_{i,j}$ is the linear shift applied to the $j$[th] basis end member in order to accurately reconstruct the $i$[th] sample diffraction pattern. Thus, peak-shifting is an additional parameter to determine during NMF runs, and is estimated via the Newton-Raphson method.

Thus, the output of shiftNMF is the end member matrix $H$, the mixing matrix $W$, and the shifting matrix $T$. We tried also shiftNMFk, developed specifically to determine the unknown number of signals[26].

While shiftNMF was designed for signal processing, realistic materials datasets requires shiftNMF (and shiftNMFk) output to obey certain physical constraints. Specifically, since the peak-shifting in diffraction patterns is caused by slight variations of the lattice parameter, there are upper bounds on the allowed peak-shifting. shiftNMF and shiftNMFk algorithms, however, require the matrix $T$ to be unconstrained. Attempts at encoding this constraint into shiftNMF has been unsuccessful so far.

Another complication arises from the increase in the overall solution space with the addition of $T$ compared to regular NMF. This can cause shiftNMF to converge to suboptimal solutions for realistic materials datasets. When a material has a correlated pair of end members or end members that share a subset of diffraction peaks, shiftNMF may blur those together.

Ultimately, our attempts to solve the problem using shiftNMF and shiftNMFk were unsuccessful. It should be noted that both methods were able to correctly recognize certain end members of materials datasets, showing promise for these methods, but incorporating physical constraints to reduce the hypothesis space should be explored in order to get physically realistic solutions.

**ACKNOWLEDGEMENTS**

The simulations were performed at the LANL in New Mexico. Research performed at Los Alamos National Laboratory was carried out under the auspices of the National Nuclear Security Administration of the United States Department of Energy. Velimir V. Vesselinov and Boian S. Alexandrov were supported by LANL LDRD grant 20180060. The work at UMD was funded by ONR N00014-13-1-0635, ONR 5289230 N000141512222, and the National Science Foundation, DMR-1505103. We are grateful to Jason Hattrick-Simpers and John Gregoire for the useful discussions and suggestions.

**CONTRIBUTIONS**

All authors designed the research. V. S., G. A., V. V., and B. A. performed the numerical calculations. V. S., B. A. and I. T. wrote the text. All participated in data analysis and editing the paper.

**COMPETING INTERESTS**

The authors declare no conflict of interest.